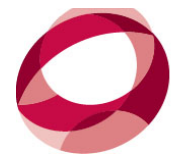

# Privacy in Information-Rich Intelligent Infrastructure

Cynthia Dwork  
Harvard University

George J. Pappas  
University of Pennsylvania

Intelligent infrastructure will critically rely on the dense instrumentation of sensors and actuators that constantly transmit streaming data to cloud-based analytics for real-time monitoring. For example, driverless cars communicate real-time location and other data to companies like Google, which aggregate regional data in order to provide real-time traffic maps. Such traffic maps can be extremely useful to the driver (for optimal travel routing), as well as to city transportation administrators for real-time accident response that can have an impact on traffic capacity. Intelligent infrastructure monitoring compromises the privacy of drivers who continuously share their location to cloud aggregators, with unpredictable consequences.[1] Without a framework for protecting the privacy of the driver's data, drivers may be very conservative about sharing their data with cloud-based analytics that will be responsible for adding the intelligence to intelligent infrastructure.

In the energy sector, the Smart Grid revolution relies critically on real-time metering of energy supply and demand with very high granularity. This is turn enables real-time demand response and creates a new energy market that can incorporate unpredictable renewable energy sources while ensuring grid stability and reliability. However, real-time streaming data captured by smart meters contain a lot of private information, such as our home activities or lack of, which can be easily inferred by anyone that has access to the smart meter data, resulting not only in loss of privacy but potentially also putting us at risk.[2]

In the previous infrastructure sectors, privacy considerations are inhibiting the sharing of data that will enable algorithms that make our transportation or energy infrastructure more intelligent. However, there are also situations where better privacy protections could enable collaboration among corporations resulting in economic growth. Infrastructure companies, such as energy traders or transportation companies (airlines), do not share data with each other because the data could reveal information about corporate strategy. This causes our infrastructure to operate far below optimal levels. Strong privacy guarantees would incentivize companies to share real-time traffic or energy data in a manner that benefits their corporate strategy, without revealing proprietary information, as well as operating our infrastructure at higher levels of operational efficiency.

To provide the intelligence in intelligent infrastructure, we will critically rely on data that is dense both spatially and temporally. The more data we have, the more intelligent our infrastructure will be, but also the more privacy may be at risk. It is hence critical to understand the tradeoff between intelligence and privacy in the context of infrastructure.

---

[1] (2015 Feb). Tracking & Hacking: Security & Privacy Gaps Put American Drivers at Risk...Retrieved May 11, 2017, from https://www.markey.senate.gov/imo/media/doc/2015-02-06_MarkeyReport-Tracking_Hacking_CarSecurity%202.pdf

[2] (2010 October). Privacy on the Smart Grid...Retrieved May 11. 2017, from http://spectrum.ieee.org/energy/the-smarter-grid/privacy-on-the-smart-grid

**State-of-the-art**

A major misconception among those concerned with privacy is that "privacy" is synonymous with "anonymity." Anonymous reporting of the data removes the identity of the data owner and releases the anonymized data. Yet the data may contain a lot of sensitive and unique information that can reveal the identity of the owner. This is problematic since "Anonymization of a data record might seem easy to implement. Unfortunately, it is increasingly easy to defeat anonymization by the very techniques that are being developed for many legitimate applications of big data." [3] In general, as the size and diversity of available data grows, the likelihood of being able to re-identify individuals (that is, re-associate their records with their names) grows substantially.[3] Indeed, the past decade has witnessed numerous re-identification attacks, where analytics applied to anonymous data, potentially coupled with additional side information, has resulted in significant privacy leaks.[4] By (erroneously) appearing to be a "have your cake and eat it too" solution, anonymity has inhibited innovation.[5]

Over the past decade, there have been major scientific breakthroughs in better understanding privacy from a scientific point of view.[6] Contrary to anonymity approaches, differential privacy has emerged as a very strong notion of privacy that allows us to think of the fundamental limits of what private information can be inferred by a malicious agent that has access to public information, or any other "side" information not derived from the database. This has led to a surge of research activity on the (differential) privacy of many basic algorithms, like learning and optimization.[7] Differential privacy has recently made its way into releases of products by both Google[8] and Apple,[9] and Uber lists familiarity with differential privacy in a job posting for Senior Privacy Engineer.[10]

While differential privacy ensures privacy of the information in data, we must consider additional privacy risks at the level of communications, as well as computations that manipulate the data. In that context, recent ideas from the cryptography community, such as homomorphic encryption and secure multi-party computation, as well as information theoretic secrecy, are promising.

The importance of privacy in the era of big data has recently resulted in a National Privacy Research Strategy.[11] Despite the centrality of privacy in the era of data driven algorithms, the national investment to address this major scientific and technological challenge. A notable significant research endeavor in this space is the DARPA BRANDEIS program, but this is an outlier. Furthermore, this white paper argues that there are unique scientific challenges that arise in the context of Intelligent Infrastructures.

---

[3] (2014 May). Big Data and Privacy: A Technological Perspective...Retrieved May 11, 2017, from https://bigdatawg.nist.gov/pdf/pcast_big_data_and_privacy_-_may_2014.pdf

[4] (2014 September) Getting to Know you... Retrieved May 11, 2017, from http://www.economist.com/news/special-report/21615871-everything-people-do-online-avidly-followed-advertisers-and-third-party

[5] (August 2009). Broken Promises of Privacy: Responding to the Surprising Failure of Anonymization… Retrieved June 1, 2017, from https://papers.ssrn.com/sol3/papers.cfm?abstract_id=1450006

[6] (2014). The Algorithmic Foundations of Differential Privacy...Retrieved May 11, 2017, from http://www.cis.upenn.edu/~aaroth/Papers/privacybook.pdf

[7] (2017). Privacy Aware Learning...Retrieved May 11, 2017, from https://stanford.edu/~jduchi/projects/DuchiJoWa12_nips.pdf

[8] (2014). RAPPOR: Randomized Aggregatable Privacy-Preserving Ordinal Response...Retrieved May 12, 2017 from https://static.googleusercontent.com/media/research.google.com/en//pubs/archive/42852.pdf

[9] (2016). What Apple's differential privacy means for your data and the future of machine learning...Retrieved May 12, 2017 from https://techcrunch.com/2016/06/14/differential-privacy/

[10] (2017). Sr. Privacy Engineer...Retrieved June 1, 2017, from https://www.uber.com/nl/careers/list/28187/

[11] (2016 June). National Privacy Research Strategy...Retrieved May 11, 2017, from https://www.nitrd.gov/PUBS/NationalPrivacyResearchStrategy.pdf

**Scientific Challenges**

It is imperative that the privacy of data information, communication, and computation be part of the national discussion about intelligent infrastructure. In this section we outline some scientific challenges regarding privacy that are unique to intelligent infrastructure.

Privacy for streaming IoT-data: A unique characteristic of IoT data that continuously monitors our infrastructure is that the data is streaming. Contrary to more static data (for example credit card data), streaming data over IoT sensors are temporally rich and continuously changing. This makes privacy over streaming data far more challenging as IoT data are temporally correlated and hence a lot of private information can be extracted by processing public information over time. This is the case, for example, with smart energy metering, where the streaming nature of the energy monitors makes it easy to recognize private activities inside our homes.

Privacy at the IoT-edge: Existing work in privacy (differential privacy and homomorphic encryption) emphasizes a cloud-based architecture view of privacy. In this architecture, IoT-devices that produce the data (by monitoring our location or energy usage) rely on a cloud aggregator that collects all data and performs computations over such data in a manner that protects their privacy. However, IoT devices have significant computational capability, and so we propose rethinking privacy from the cloud to the edge devices of an IoT architecture. Providing privacy at the level of the sensor before transmitting the data to the cloud eliminates the need to place trust in a central aggregator of data. This is the philosophy behind the "local model" of differential privacy, currently employed by Google and Apple. Encrypting the data using homomorphic encryption allows computation on the original data, although it does not provide the privacy guarantees of differential privacy.

Decentralized Private Computation: Collecting data in a single cloud creates a privacy risk through accidental disclosures or malicious attacks to a centralized third-party data aggregator. Even encrypted data can be subject to legal coercion. Centralized systems also suffer reliability concerns. Secure Multiparty Computation provides an alternative, and combines well with differential privacy.[12] In such decentralized approaches, it would be possible to compute across many IoT-devices even if the data would reside physically on the IoT devices. Such decentralized approaches would also enable collaboration across corporations (say energy traders) that would like to perform real-time energy bidding without sharing critical data that may compromise either private data or proprietary strategy.

Variable privacy: Current privacy approaches provide static guarantees across time, space, or even data owners. However, the data collected by monitoring infrastructure, may have variable levels of privacy that change across time (security may require us to lower or raise privacy over time), across space (some areas may contain much more sensitive information than others), data (some data may be less sensitive than others), or users (some users may be willing to share their data more than others). Given the spatial and temporal nature of IoT data, it is important that develop privacy principles that can adjust the level of guarantees accordingly.

---

[12] (2006). Our Data, Ourselves: Privacy Via Distributed Noise Generation...Retrieved June 1, 2017, from http://www.wisdom.weizmann.ac.il/~naor/PAPERS/odo.pdf

Event-based privacy: Existing privacy approaches ensure the privacy of the data or the computation with private data. But in many situations, there are weaker requirements on privacy and that is protecting the privacy of spatio-temporal events. For example, while monitoring our house using smart meters, we may decide that we may wish to protect the privacy of critical private events, such as taking a bath.[13] How to we develop an approach for privacy that hides private events while releasing information about our energy usage at home that is critical for demand response? Alternatively, how can we create privacy mechanisms that detect safety-critical events (such as detecting fire, or someone asking for help) without recording or processing other private information that is not critical?

We have already discussed the limitations of de-identification, but what about purely statistical data analysis? Unfortunately, this too has its limits. A useful analogy is to the telling and retelling of a story, with random details altered each time to protect privacy of the principals. If all the versions are collected and collated, the obscured details emerge. Similarly, if data are sliced and diced to a significant degree, then details of the individual data records can often be reconstructed. These problems persist even if the statistics are perturbed with the goal of providing privacy for the data subjects. A host of mathematical results tell us that "overly accurate" estimates of "too many" statistics can completely destroy privacy, a potentially counterintuitive phenomenon colloquially known as the *Fundamental Law of Information Recovery*.

Consequently, unless we abandon privacy, we must prioritize even among privacy-preserving data usages.

**Actions and Recommendations**

We therefore recommend the following actions:

1. **Infrastructure Data:** One of the main challenges in developing privacy-aware data analytics is access to data. In this context, we recommend the development of a depository for IoT-data that is monitoring different infrastructure sectors (transportation, energy, water, etc.). Such depositories will be crucial for the development as well as testing of privacy-aware analytics.
2. **Joint Funding Initiative:** Develop a joint interagency research program across relevant agencies (NSF, DoT, DoE, DHS) where fundamental scientific advances are funded by NSF and the contextualization of such principles and algorithms are funded by relevant sectors (DoE for SmartGrid Privacy, DoT for transportation privacy, etc.). The operational model for such a interagency program can be similar to the National Robotics Initiative (NRI) or the Cyber-Physical Systems (CPS) program.
3. **The National Epsilon Registry**: Differential privacy provides a measure of privacy loss, typically called $\epsilon$ ("epsilon"), and differentially private algorithms are parameterized, so that the total privacy loss incurred by running the algorithm can be bounded by any desired value $\epsilon$. Smaller $\epsilon$ means better privacy, but typically also results in lower utility. Theoretical work shows how privacy losses combine as data are used and re-used (in the worst case, the epsilons add up). Dwork and Mulligan have proposed an "Epsilon Registry", in analogy to a toxic release registry, in which firms and websites that traffic in personal information would record details about their treatment of data. Typical entries in the registry would record, for example: (1) pathways of privacy loss (what is being done with the data that leads to privacy loss?); (2) granularity (is a single search by the user being protected at a given level $\epsilon$ of privacy loss, or is the set of all searches by this user simultaneously being protected at this level?); (3) what is the value of epsilon being used per datum; (4) burn-rate (is the privacy loss limited to $\epsilon$ over the lifetime of the data? Per day? Per week?); (5) what is the cumulative privacy loss incurred before data are retired?; and (6) technical variant of differential privacy used. In the spirit of "sunlight is the best disinfectant," it is hoped that revealing this information will inspire competition among firms, bringing the domain-specific knowledge of those who profit from the data to bear on its safe deployment.

---

[13] (2001). "The Agema Thermovision 210 might disclose, for example, at what hour each night the lady of the house takes her daily sauna and bath..." Justice Antonin Scalia from KYLLO V. UNITED STATES (99-8508) 2001, available here: https://www.law.cornell.edu/supct/html/99-8508.ZO.html.

4. **Data Property Rights:** Defining property rights over data and information will be very important both in protecting data owners and also in creating a new economy for data. For example, when I purchase a book from amazon, by default Amazon has rights over that record equivalent to ownership. It obtains it by bundling the sale with the terms of service. There is no option to buy the book without relinquishing rights over the record of the transaction. If property rights could be defined then there is a possibility for them to be traded. Amazon, could for example, pay me for the data I share with them. As we move forward in instrumenting our works with billions of sensors owned by many owners, having well defined property rights could be extremely important as well as enabling the new data economy.

5. **Privacy Forum:** Privacy requires a continuing discussion among regulators, legal experts, philosophers, privacy technology experts, and corporations, in order to balance the scientific feasibility of privacy with social norms of privacy. Regulators need to be informed about the scientific limits of privacy as they consider updating regulations, and vice versa, technologists needs to be aware of where technology ends and where the law begins in this space. Corporations and innovators need such a forum to ensure regulation that provides meaningful privacy guarantees while enabling innovation as much as possible. This requires more frequent communication among sectors of society that do not often meet. We propose having an annual Privacy Forum, where technologists, computer scientists, IoT engineers, legal experts, companies, regulators, and legal experts convene to discuss the moving interfaces among these disciplines as it applies to IoT privacy.

**Acknowledgements**

Thanks to Rakesh Vohra, Aaron Roth, and Andreas Haeberlen at the University of Pennsylvania for discussions on the privacy of IoT data, and to Helen Wright for invaluable help in writing and editing.

*This material is based upon work supported by the National Science Foundation under Grant No. 1136993. Any opinions, findings, and conclusions or recommendations expressed in this material are those of the authors and do not necessarily reflect the views of the National Science Foundation.*